\documentclass[twocolumn,amssymb,prd,amssymb]{revtex4}
\usepackage{bm}
\newcommand{\be}{\begin{equation}}
\newcommand{\ee}{\end{equation}}
\newcommand{\ba}{\begin{eqnarray}}
\newcommand{\ea}{\end{eqnarray}}
\begin{document}
\title{Time-Varying Cosmological Term: Emergence \\
and Fate of a FRW Universe}

\author{R. Aldrovandi}
\email{ra@ift.unesp.br}

\author{J. P. Beltr\'an Almeida}
\email{jalmeida@ift.unesp.br}

\author{J. G. Pereira}
\email{jpereira@ift.unesp.br}

\affiliation{Instituto de F\'{\i}sica Te\'orica,
Universidade Estadual Paulista \\
Rua Pamplona 145, 01405-900 S\~ao Paulo SP, Brazil}

\begin{abstract}
A time-varying cosmological ``constant'' $\Lambda$ is consistent with Einstein's
equation, provided matter and/or radiation is created or destroyed to compensate for
it. Supposing an empty primordial universe endowed with a very large  cosmological
term, matter will emerge gradually as $\Lambda$ decays. Provided only  radiation or
ultrarelativistic matter is initially created, the universe starts in  a nearly de
Sitter phase, which evolves towards a FRW r\'egime as expansion  proceeds. If, at some
cosmological time, the cosmological term begins increasing  again, as presently
observed, expansion will accelerate and matter and/or radiation will be transformed
back into dark energy. It is shown that such accelerated expansion is a route towards
a new kind of  gravitational singular state, characterized by an empty, conformally
transitive spacetime in which all energy is dark. 
\end{abstract}


\maketitle

\renewcommand{\thesection}{\arabic{section}}
\section{Introduction}

One of the fundamental problems of cosmology is whether the universe  will someday
re-collapse in a big crunch, or will expand forever  becoming increasingly cold
and empty. Recent cosmological  observations involving both supernovae \cite{sn}
and the cosmic  microwave background \cite{cmb} suggest that the universe 
expansion is accelerating, an effect which points to the presence of some kind  of
negative-pressure energy, generically called dark energy. This  energy is usually
described by an equation of state of the form $\omega=p/\rho $, where $\omega$ is
a parameter, not necessarily  constant, and $p$ and $\rho$ are respectively the
dark energy  pressure and density. Cosmic acceleration requires that $\omega < 
-1/3$. The simplest explanation for dark energy is a cosmological  term $\Lambda$,
for which $\omega = - 1$. Other popular, though  somewhat bizarre possibilities
are quintessence \cite{5essence}, a  cosmic scalar field in which $-1< \omega <
-1/3$, and phantom energy \cite{phantom}, scalar field models presenting a quite
unusual  kinetic term, for which $\omega < -1$. Depending on both the model  and
the value of the parameters, different fates for the universe can  be achieved,
which range from a simple re-collapse, passing through a  bleak eternal expansion,
to an astonishing big rip end \cite{br}.

Working in the context of a {\em dynamical cosmological term}, a different and
unexplored possibility for the fate of the universe will be presented.  We begin
in section 2 where, for the sake of completeness, we show that a dynamical
cosmological term is consistent with general relativity, provided matter and/or
radiation is created to make overall energy conserved. In section 3, the Friedmann
equations for a time-depending $\Lambda$ are obtained, and in section 4 a
qualitative analysis of such evolution equations is made for the specific case of a
{\it time-decaying} $\Lambda$. In spite of the lacking of an adequate understanding of
the physics associated with a cosmological term, in particular of the law governing its
time evolution, it is pointed out that an appropriate $\Lambda$-evolution could explain
most of the present-day observations, without necessity of any exotic machinery. In
order to comply with the present observational data, we consider in section 5 the case
of an increasing $\Lambda$, which will produce an accelerated universe expansion. It is
then pointed out that this accelerated expansion corresponds to a new route to a
collapsing universe, whose final stage is an empty, singular, conformally
transitive spacetime in which all energy is in the form of dark energy 
\cite{cone}.

\section{Dynamic dark energy and Einstein's equation}

In the presence of a cosmological constant $\Lambda$, Einstein's equation assumes
the form \cite{sign}
\be
G^\mu{}_\nu \equiv R^\mu{}_\nu - \frac{1}{2} \, \delta^\mu{}_\nu R =
\frac{8 \pi G}{c^4} \, \left[T^\mu{}_\nu + \frac{c^4 \Lambda}{8 \pi 
G} \,
\delta^\mu{}_\nu \right],
\label{einstein}
\ee
where $T^\mu{}_\nu$ is the energy-momentum density tensor of the 
source field. Put together,
the Bianchi identity
\be
\nabla_\mu G^\mu{}_\nu = 0
\label{bianchi}
\ee
and the source ``covariant conservation law''
\be
\nabla_\mu T^\mu{}_\nu = 0 
\label{colaw0}
\ee
imply that the cosmological constant cannot present any kind of space or time
dependence. In other words, it must be a true constant. On the other hand, since
inflationary models require a very high $\Lambda$ at the early stages of the
universe \cite{inflation} and present-day observations indicate a much smaller
value \cite{sn,cmb}, that constancy restriction appears as one of the central
problems of cosmology \cite{cosmocon}.

Of course, the vanishing of a (gravitational field dependent) covariant divergence
as in (\ref{colaw0}) is no true conservation law: it yields no time-conserved
``charge''.  The role of such a ``covariant conservation'' is to regulate the
exchange of energy and momentum between the source fields (generically called
matter, from now on) and the gravitational field.  Furthermore, it is not
necessarily true in all circumstances.  It would not be expected to hold, for
example, if matter (plus the gravitational field it gives rise to) is being {\em
created} from an independent source.  As already pointed out in the literature
\cite{vish1}, a time-decaying cosmological term can be such a source.

We observe to begin with that, in the presence of a non-constant cosmological
term, what is imposed by the use of the Bianchi identity (\ref{bianchi}) in
Einstein's equation (\ref{einstein}) is, instead of (\ref{colaw0}), the condition
\be
\nabla_\mu \left[T^\mu{}_\nu +
\Lambda^\mu{}_\nu \right] = 0,
\label{colaw1}
\ee
where $\Lambda^\mu{}_\nu = \varepsilon_\Lambda \delta^\mu{}_\nu$
is the dark energy-momentum tensor associated with the cosmo\-logical
term, with
\be \varepsilon_\Lambda = \frac{c^4 \Lambda}{8 \pi G}
\label{deden} 
\ee 
the corresponding energy density.  The energy-momen\-tum of matter alone is
conse\-quently not covariantly conserved.  Only its sum with the dark
energy-momen\-tum tensor is.  The covariant conservation law (\ref{colaw1}) can be
interpreted as a constraint regulating the exchange of energy and momentum between
matter, gravitation and the cosmological term.  In other words, it says how the
cosmological dark energy can be transformed into ordinary matter plus the
gravitational field it engenders, or vice-versa.  Assuming that $\Lambda$ depends
only on the cosmological time $t$ \cite{spdep}, the covariant conservation
(\ref{colaw1}) is equivalent to (we use $i, j, k = 1, 2, 3$ to denote space indices)
\be
\nabla_\mu
T^\mu{}_i = 0,
\label{colaw2}
\ee
and
\be
\nabla_\mu T^\mu{}_0 = - \frac{c^3}{8 \pi G} \, \frac{d \Lambda}{dt}.
\label{colaw3}
\ee

We see now from Eq.~(\ref{colaw3}) that a time-decaying $\Lambda$ implies that the
source energy-momentum tensor is not {\it covariantly} conserved, and consequently
matter must necessarily be created as the cosmological term decays.  Notice that
the total energy of the universe is conserved despite matter creation.  To see
that, it is enough to take Einstein's equation with a cosmological term in the so
called potential form \cite{moller}, \be \partial_\rho (\sqrt{-g}
\, {\mathcal S}^{\rho \mu}{}_\nu) = \frac{8 \pi G}{c^4} \left[
\sqrt{-g} (t^\mu{}_\nu + T^\mu{}_\nu + \Lambda^\mu{}_\nu) \right], \ee where
${\mathcal S}^{\rho \mu}{}_\nu = - {\mathcal S}^{\mu \rho}{}_\nu$ is the
superpotential, and $t^\mu{}_\nu$ is the energy-momentum pseudo\-tensor of the
gravitational field.  Due to the anti-symmetry of the superpotential in the first
two indices, the total energy-momentum density, which includes the gravitational,
the matter and the cosmological parts, is conserved:
\be
\partial_\mu
\left[ \sqrt{-g} (t^\mu{}_\nu + T^\mu{}_\nu + \Lambda^\mu{}_\nu)
\right] = 0.
\label{noether}
\ee
This is actually the Noether conservation law obtained from the 
invariance of the theory under a
general transformation of the spacetime coordinates.

It is important to remark that the covariant conservation law (\ref{colaw3}) is
different from that appearing in quintessence models \cite{5essence}.  In fact,
in such models the energy-momen\-tum tensor of the scalar field that replaces the
cosmological term is itself covariantly conserved, and consequently there is no
matter creation.  On the other hand, despite the presence of continuous matter
creation, this mechanism is different also from the $C$-field theory of Hoyle and
Narlikar \cite{HN} as in the present model no scalar field is introduced, but only
a cosmological term whose time decaying turns out to be linked to the matter
energy density evolution through the Friedmann equations.

\section{Friedmann equations}

The starting point of our considerations will be an empty universe endowed with a
very large---possibly infinite \cite{infi}---decaying cosmological term.  In the
extreme case of an infinite $\Lambda$, this spacetime is given by a singular
cone-space, transitive under proper conformal transformations \cite{cone}.  If we
assume that the ensuing newly created matter is a homo\-geneous and isotropic
fluid, it is natural to consider that the metric tensor of this gravitational
field be of the Friedmann-Robertson-Walker (FRW) type
\[
ds^2 = c^2 dt^2 - a^2
\left[ \frac{dr^2}{1 - k r^2} + r^2 (d\theta^2 + \sin^2\theta d\phi^2)
\right],
\]
where $a=a(t)$ is the expansion factor and $k$ is the
curvature parameter of the space section.  The coordinates are taken
for a comoving observer in relation to the lines of flux of a perfect
fluid, whose energy-momentum tensor has the form
\be
T^\mu{}_\nu =
(\varepsilon_m + p_m) u^\mu u_\nu - p_m \delta^\mu{}_\nu,
\ee
with $p_m$ and $\varepsilon_m$ the pressure and the energy density of the
created matter.  Denoting $x^0 = c t$, $x^1 = r$, $x^2 = \theta$, and
$x^3 = \phi$, the non-zero energy-momentum components for that
homogeneous, isotropic fluid will be:
\[
T^1{}_1 = T^2{}_2 = T^3{}_3 = -\, p_m,
\]
\[
T^0{}_0 = \varepsilon_m.
\]
The conservation law (\ref{colaw3}) in this case becomes \be \frac{d
\varepsilon_m}{dt} + 3 H (\varepsilon_m + p_m) = -\, \frac{d
\varepsilon_\Lambda}{dt},
\label{colaw4}
\ee
with
\[
H = \frac{1}{a} \frac{da}{dt}
\]
the Hubble parameter.  This is actually one of the Friedmann equations.  In fact,
it can be seen that it follows from the combination of the usual Friedmann
equations
\begin{equation}
\left(\frac{da}{dt} \right)^2 = \left[ \frac{8 \pi G}{3 c^2} \,
\varepsilon_m + \frac{\Lambda c^2}{3} \right] a^2 - k c^2
\label{fri1}
\end{equation}
and 
\be
\frac{d^2a}{dt^2} = \left[\frac{\Lambda c^2}{3} - \frac{4 \pi
G}{3 c^2} \left(\varepsilon_m + 3 p_m \right) \right] a,
\label{fri2}
\ee 
provided $\Lambda$ is time dependent. It is important to remark that, according to
this model, matter is not created at once in a big bang. It emerges as long as the
cosmological term decays, in a gradual process.

Let us now suppose that the newly created ordinary matter satisfies 
an equation of state of the form
\be
p_m = \omega_m \,  \varepsilon_m,
\label{baro}
\ee
with $0\leq\omega_m\leq 1$  a parameter that depends on the specific  kind of
matter. Of course, the matter content of the universe can be made up of more than
one  component, each one satisfying an equation of state with a different 
$\omega_m$. These possibilities should be taken into account in a  comprehensive
description of the universe evolution. Here, however,  it is enough for our
purposes to consider a one-component. In this  case, Eq.~(\ref{colaw4}) becomes
\be
\frac{d \varepsilon_m}{dt} + 3 H (1+\omega_m) \varepsilon_m = - 
\frac{d \varepsilon_\Lambda}{dt}.
\label{colaw6}
\ee
The second Friedmann equation, on the other hand, can be written in 
the form
\begin{equation}
\frac{d^2a}{dt^2} = \frac{8 \pi G}{3 c^2} \left[\varepsilon_\Lambda - 
\frac{1}{2}
\left(1 + 3 \omega_m  \right)\varepsilon_m \right] a. 
\label{sfe}
\end{equation}

\section{Evolution analysis}
 
\subsection{General Case}

The Friedmann equation (\ref{colaw6}) establishes a connection between
the evolutions of $\varepsilon_m$ and $\varepsilon_\Lambda$.  In fact,
if $\varepsilon_m$ depends on time through the expansion factor $a$,
the dark energy density $\varepsilon_\Lambda$ will also have the same
dependence on $a$, and vice-versa.  In principle any behavior is
possible for $\varepsilon_m$ and $\varepsilon_\Lambda$, although it is
usual to suppose that $\varepsilon_m$ evolves as a power law in the
expansion factor,
\be
\varepsilon_m = \alpha \, a^{-n},
\label{maden}
\ee
with $\alpha$ a constant and $n$ a number (integer or not)
\cite{hands}. In this case, Eq.~(\ref{colaw6}) implies 
\be
\varepsilon_\Lambda = \frac{3 (1+\omega_m ) - n}{n} \; \varepsilon_m,
\label{epstoeps}
\ee 
where we have assumed a vanishing integration constant.  In the presence of a
dynamical cosmological term, therefore, depending on the parameters $n$ and
$\omega_m$, the energy densities $\varepsilon_m$ and $\varepsilon_\Lambda$ may
eventually be of the same order, as strongly suggested by present observations
\cite{sean}.  Of course, these parameters can also lead to periods in which
$\varepsilon_m$ and $\varepsilon_\Lambda$ are completely different. It is
interesting to observe that the case $n=0$, which would correspond to an
equilibrium between matter creation and universe expansion ($\varepsilon_m$ =
constant), is excluded by the Friedmann equations.  Notice furthermore that, for
$\varepsilon_\Lambda$ constant, Eq.~(\ref{colaw6}) yields the solution
$\varepsilon_m \sim a^{-3(1+\omega_m)}$.  For a time-decaying
$\varepsilon_\Lambda$, however, $n$ is required to be in the interval
\be
0 < n < 3 (1+\omega_m ).
\label{nrange}
\ee
Since matter is continuously created, it is natural that
$\varepsilon_m$ evolves at a rate slower than $a^{-3(1+\omega_m)}$, which would be
its behavior if matter were not being created.

On the other hand, using the equation of state (\ref{baro}), as well
as the relations (\ref{maden}) and (\ref{epstoeps}), the Friedmann
equation (\ref{sfe}) becomes
\begin{equation}
\frac{d^2a}{dt^2} = \frac{3 (1+\omega_m ) \beta^2}{2} 
\left(\frac{2-n}{n} \right) a^{1-n},
\end{equation}
where $\beta^2=8 \pi G \alpha/3c^2$.  We see from this equation that, for $n=2$,
$n>2$ and $n<2$, the universe expansion acceleration will be respectively zero,
negative and positive.  This property could eventually explain why the
acceleration was negative in the past and positive today, as suggested by recent
observational data \cite{sean}.  Furthermore, in the case of a positive
acceleration ($n<2$), the ranges $n>1$ and $n<1$ will represent respectively the
cases in which the acceleration is decreasing or increasing, with the value $n=1$
representing a universe with a constant expansion acceleration, given by
\[
\frac{d^2a}{dt^2} = \frac{3 (1+\omega_m ) \beta^2}{2}.
\]
In this case, $a \sim t^2$, and we have the relations
\[
\Lambda \sim a^{-1} \sim H^2 \sim t^{-2}.
\]
We notice finally that, as the parameters $\omega_m $ and $n$ have  very limited
ranges, the above results do not change very much when $\omega_m $ is assumed to
vary  slowly with the cosmological time, or the matter content of the universe has
more than one  component.

\subsection{The Flat Case}

Recent observational data favor a universe with flat spatial sections ($k=0$). In
this case, it is possible to find an explicit time-dependence for the cosmological
term, which is valid for any value of the parameters $n$ and $\omega_m$. In fact,
for $k=0$ the Friedmann equation (\ref{fri1}) can be written in the form
\begin{equation}
\left(\frac{da}{dt} \right)^2 = \frac{3 \beta^2 (1+\omega_m)}{n} \, 
a^{2-n},
\end{equation}
or equivalently
\begin{equation}
a^{n/2 -1} \; da = \left(\frac{3(1+ \omega_m) \beta^2}{n} 
\right)^{1/2} \, dt.
\label{frik0}
\end{equation}
Assuming a vanishing integration constant \cite{foot4}, the solution is found to be
\begin{equation}
a = \left( \frac{3 n(1+ \omega_m) \beta^2}{4}\right)^{2/n} \; t^{2/n}.
\end{equation}
As a consequence, the matter and the dark energy densities will  present the
behavior \cite{foot5}:
\begin{equation}
\varepsilon_m \sim \varepsilon_\Lambda \sim t^{-2}.
\end{equation}
Due to relation (\ref{deden}), and using Einstein's equation, we  have also
\begin{equation}
\Lambda \sim R \sim t^{-2}.
\label{k0behavior}
\end{equation}
This is the time-dependence of $\Lambda$ and $R$ for any value of the parameters
$n$ and $\omega_m$.  We see from this behavior that both the cosmological term
$\Lambda$ and the scalar curvature $R$ diverge at the initial time, which signals
the existence of an initial singularity.

As an example related to the initial period of the universe, let us assume what
can be called a {\it fiat lux} hypothesis, according to which the newly created
matter satisfies the ultra-relativistic equation of state \cite{weinberg}
\be
\varepsilon_m = 3\, p_m,
\label{radiation}
\ee
which corresponds to $\omega_m =1/3$. In this case, the trace of Einstein's
equation (\ref{einstein}) gives \cite{foot6}
\be
R = -\, 4 \,
\Lambda,
\label{r4lambda}
\ee 
where we have used $T \equiv T^\mu{}_\mu = \varepsilon_m - 3 p_m = 0$.  As long as
only radiation and ultrarelativistic matter is created, the scalar curvature is
completely determined by $\Lambda$, and in this sense the universe can be
considered to be in a nearly de Sitter phase (of course, since the cosmological
term is not constant, it is not a de Sitter spacetime in the ordinary sense). 
Now, for a positive cosmological term $(\Lambda > 0)$, the scalar curvature is
given by $R = - 12/L^2$, where $L$ is the de Sitter length-parameter, or de Sitter
``radius''.  In terms of $L$, the dark energy density (\ref{deden}) becomes
\be
\varepsilon_\Lambda = \frac{3 c^4}{8 \pi G
L^2}.
\label{dedenL} 
\ee
On the other hand, the Friedmann equation (\ref{colaw6}) assumes the  form
\be
\frac{d \varepsilon_m}{dt} + 4 H \varepsilon_m =
-\, \frac{d\varepsilon_\Lambda}{dt}.
\label{f1}
\ee
For $\varepsilon_\Lambda$ constant, it yields the usual solution 
$\varepsilon_m \sim a^{-4}$. For a time-decaying $\varepsilon_\Lambda$, however,
we get the  relation
\be
\varepsilon_\Lambda = \frac{4 - n}{n} \; \varepsilon_m,
\label{darkma}
\ee
where now $0<n<4$. Equations (\ref{dedenL}) and (\ref{darkma}) imply  that, as
long as matter is ultrarelativistic, the de Sitter radius expands according to
\be
L^2 = \frac{c^2}{3 \beta^2} \left(\frac{n}{4 - n}\right) \, a^{n}.
\ee

\section{Final remarks}

In the context of a $\Lambda$ cosmology, we can say that the dynamics  of the
universe was dominated by a very large positive cosmological term during the 
period of primordial inflation, or even by the eventual extreme possibility of an
infinite $\Lambda$ \cite{infi}. Quantum fluctuations could then give rise
to a de Sitter spacetime, which is well known to exhibit a horizon \cite{paddy}
at the de Sitter length $L=\sqrt{3/\Lambda}$. If, at the Planck time, $L$ is
assumed to coincide with the Planck length $l_P$, the cosmological ``constant'' 
would, at that moment, have the value
\[
\Lambda = 3 / (l_P)^2 \simeq 1.2 \times 10^{66}\, {\rm cm}^{-2}.
\label{vacuum}
\]
The dark energy density, on the other hand, would be
\[
\varepsilon_{\Lambda} \simeq 10^{112}\,{\rm erg/cm^3}.
\]
At this time, therefore, most of the energy density of the universe  would be in
the dark energy form. We notice in passing that, since present-day observations 
indicate that
\[
\varepsilon_{\Lambda}^{0} \simeq 10^{-8}\,{\rm erg/cm^3},
\]
the evolving mechanism implied by the Friedmann equations with a  time-decaying
cosmological term could eventually give an account of  this difference. As the
$\Lambda$ term decays and the universe  expands, matter and/or radiation is
gradually created, giving rise to  a FRW universe. Despite the continuous creation
of matter, however,  the total energy density of the universe---which includes, in 
addition to the matter and the dark energy densities, the energy  density of the
evolving gravitational field they generate---is  conserved. We remark once more
that, according to this mechanism,  matter is not created at once in a big bang,
but gradually as the  cosmological term decays.

Models with a time decaying cosmological term \cite{decay} have  already been
extensively considered in the literature \cite{vish3}. The basic idea underlying 
these models is to try to explain how a large primordial $\Lambda$ can present a
small value  today. All of them are essentially phenomenological in nature, and
based preponderantly on  dimensional arguments. Here, however, instead of adopting
a  phenomenological point of view, we have followed a theoretical  approach based
almost exclusively on the equations governing the  universe evolution, that is, on
the Friedmann and on the matter  equations of state. Since a new degree of
freedom, connected with the  time-evolving cosmological term, is introduced, there
remains in the  theory a free parameter---the cosmological term---whose time 
evolution has eventually to be determined by further fundamental 
physics \cite{foot7}. Of course, to explain  why $\varepsilon_{\Lambda}$ and
$\varepsilon_m$ are approximately of  the same order today, and eventually why the
acceleration was  negative in the past and positive today, as strongly suggested
by  recent observations, a quite specific time evolution for $\Lambda$ is
necessary.  This question, however, remains as one of the mysteries involving the 
nature of dark energy, an open problem to be investigated. The  important point is
to observe that a dynamical cosmological term  endowed with an appropriated time
evolution contains enough free  parameters to allow a wide range of scenarios for
the cosmological  evolution, including the main features favored by recent
astronomical  data, and does not require any further exotic ingredient (as, for 
example, phantom energy, quintessence models, or even modifications  of the
gravitational theory) to consistently describe the dynamical  evolution of the
universe \cite{horvat}.

An important point is to observe that, in order to allow the formation of the
cosmological structures we see today (galaxies,  clusters of galaxies, and so on), the
universe necessarily has passed  through a period of non-accelerating expansion, which
means that the  cosmological term must have assumed a tiny value during some 
cosmological period in the past. On the other hand, recent observations \cite{sn,cmb}
indicate that the universe is presently entering  another exponential expansion era.
Even though we still lack an adequate understanding of the basic physics associated
with the evolution of the  cosmological term, the above facts put together suggest a
primordial universe  characterized by a very large $\Lambda$, including eventually
the  possibility of an infinite $\Lambda$, followed by a somehow decaying 
cosmological term which, after keeping a minimum value during some period, has entered
a new increasing period. If this is true, in the  same way a decaying $\Lambda$
implies that matter and/or radiation be  created, conservation of energy requires that
matter and/or radiation  be transformed into dark energy by a time increasing
$\Lambda$ term \cite{foot8}.

Now, it is frequently argued that, if this new phase of exponential expansion is in
fact occurring, the universe would be driven either to a bleak future, a state that
could be called cosmic loneliness, or eventually to a complete disintegration  which
has been called ``big rip'' \cite{br}. However, as far as a time increasing
$\Lambda$-term implies that matter and/or radiation be transformed into dark energy,
such a mechanism could eventually lead the universe to a state in  which the whole
energy would be in the form of dark energy. In other words, a time increasing
$\Lambda$ does not necessarily mean that the universe will disperse and become colder,
but that it may be moving towards a new kind of singular state. If led to the extreme
situation of an infinite cosmological term, the universe would achieve a singular
state characterized by an empty, causally disconnected, conformally transitive
spacetime (a brief description of the basic geometrical properties of this spacetime
is given in the Appendix). Of course, whether quantum effects will or not preclude
such ``collapse'' is an open question.

\begin{acknowledgments}
The authors would like to thank S. Carneiro, J. V. Narli\-kar and J. Overduin for useful
comments. They would like also to thank FAPESP, CAPES, and CNPq for partial financial
support.
\end{acknowledgments}

\begin{appendix}
\section*{Appendix: The infinite-$\Lambda$ spacetime}

\subsection*{A.1~~Kinematic groups: transitivity}

The kinematic group of any spacetime will always have a subgroup accounting for  both
the isotropy of space (rotation group) and the equivalence of inertial frames
(boosts). The  remaining transformations, which can be either commutative or not, are 
responsible for the homogeneity of space and time. This holds, of course, for usual
Galilean and other  conceivable non-relativistic kinematics \cite{levy}, but also for
special-relativistic  kinematics. The best known relativistic example is the
Poin\-ca\-r\'e group ${\mathcal P}$, naturally  associated with the Minkowski
spacetime $M$ as its group of motions. It contains, in the form of a  semi-direct
product, the Lorentz group ${\mathcal L} = SO(3,1)$ and the translation group
${\mathcal T}$.  The latter acts transitively on $M$ and its manifold is just $M$.
Indeed, Minkowski  spacetime is  a homogeneous space under ${\mathcal P}$, actually
the quotient $M \equiv {\mathcal T} = {\mathcal  P}/{\mathcal L}$. The invariance of
$M$ under the transformations of ${\mathcal P}$ reflects its  uniformity. The Lorentz
subgroup provides an isotropy around a given point of $M$, and translation invariance
enforces this isotropy around any other point.  This is the usual meaning of
``uniformity", in  which
${\mathcal T}$ is responsible for the equivalence of all points of spacetime.

\subsection*{A.2~~The case of the de Sitter spacetime}

The de Sitter $dS(4,1)$ and anti-de Sitter $dS(3,2)$ spacetimes are the only 
possible uniformly {\em curved} four-dimensional metric spacetimes \cite{weinberg}.
They are maxi\-mally--sym\-me\-tric, in the sense that they can lodge the maximum
number of Killing vectors. These spacetimes  are related respectively to a positive
and to a negative cosmological term $\Lambda$, and  their groups of motions are
respectively the de Sitter $SO(4,1)$ and anti-de Sitter $SO(3,2)$ groups. Both
spaces are homogeneous \cite{livro}:
\[
dS(4,1) = SO(4,1)/ SO(3,1),
\]
\[
dS(3,2) = SO(3,2)/ SO(3,1).
\]
In addition, each group manifold is a bundle with the corresponding de Sitter or 
anti-de Sitter space as base space, and the Lorentz group ${\mathcal L}$ as fiber
\cite{kono}.

Let us then analyze the kinematic group of the de Sitter spacetime $(\Lambda>0)$. In
terms of the stereographic coordinates $x^a$ ($a, b, \dots=0,1,2,3$), the generators
of infinitesimal de Sitter transformations are written as \cite{gursey}
\be
J_{a b} =
\eta_{ac} \, x^c \, P_b - \eta_{bc} \, x^c \, P_a
\label{dslore}
\ee
and
\be
T_{a} = \Big(L \, P_a - \frac{1}{4 L} \, K_a \Big),
\label{dstra}
\ee
where
\be
P_a = \frac{\partial}{\partial x^a} \quad \mbox{and} \quad
K_a = \left(2 \eta_{ab} x^b x^c - \sigma^2 \delta_{a}{}^{c}
\right) P_c
\label{cp2} 
\ee
are, respectively, the generators of translations and {\it proper} conformal 
transformations. In the above expressions, $L = (3/\Lambda)^{1/2}$ is a
length-parameter related to the curvature of the de Sitter space, and $\sigma^2 =
\eta_{ab} x^a x^b$ is the Lorentz invariant interval, with $\eta_{ab} =
\mbox{diag}(+1,-1,-1,-1)$. The generators $J_{a b}$ refer to the Lorentz subgroup
${\mathcal L}$, whereas $T_{a}$ define the transitivity on the corresponding 
homogeneous space. According to Eq.~(\ref{dstra}), we see that the de Sitter
spacetime is transitive under a mixture of translations and proper conformal
transformations. The relative importance of each one of these transformations is
determined by the value of the cosmological term.

\subsection*{A.3~~Contraction limits}

Let us begin by remarking that, on account of the quotient character of the de  Sitter
spacetime, geometry and algebra turns out to be deeply connected: any deformation in
the algebras and  groups will produce concomitant deformations in the imbedded
spacetime. As an example, let us  consider first the limit $\Lambda \to 0$ (which
corresponds to $L \to \infty$). In this limit, as is well known \cite{gursey}, the
de Sitter group is contracted \cite{foot9} to the Poincar\'e group ${\mathcal P} =
{\mathcal L} \, \oslash \, {\mathcal T}$. This group deformation will produce changes
in the imbedded  spacetime. In fact, the de Sitter spacetime reduces in this limit to the
flat Minkowski space $M = {\mathcal P}/{\mathcal L}$, which is {\em transitive}  under
translations only.

In the limit  $\Lambda \to \infty$ (which corresponds to $L \to 0$),  the de Sitter
group is contracted to the so called {\it second} or {\it conformal}  Poincar\'e group
${\mathcal Q}$, the semi-direct product between Lorentz ${\mathcal L}$ and the proper
conformal group ${\mathcal C}$, that is, ${\mathcal Q} = {\mathcal L} \, \oslash \,
{\mathcal C}$ \cite{ap1}. This group deformation will accordingly produce changes in
the imbedded spacetime. In fact, in the limit of an infinite cosmological term, the de
Sitter  space is led to a four-dimensional cone-space \cite{cone}, which we denote by
$N$.  Like Minkowski, the cone-space $N$ is a homogeneous space, but under ${\mathcal 
Q}$: $N = {\mathcal Q}/{\mathcal L}$. The kinematical group ${\mathcal Q}$, as the 
Poincar\'e group, has the Lorentz group ${\mathcal L}$ as the subgroup accounting for
the isotropy of $N$. However, the proper conformal transformations introduce a new
kind of homogeneity: instead of  the ordinary translations, which defines the
homogeneity on Minkowski spacetime, all points of $N$ are equivalent through proper
conformal transformations. In other words, the cone-space $N$ is  transitive under
proper conformal transformations. On account of this conformal transitivity, the
cone-space $N$ can be said to be {\em conformally infinite}.

It is important to remark that the usual metric of the de Sitter spacetime becomes
singular in the contraction process \cite{cone}. This is the reason why the  ordinary
notions of space distance and time interval fail to exist on $N$. However, the
corresponding notions of {\it  conformal space} and {\it conformal time} can be
defined through the introduction  of the conformal invariant metric \cite{cone}
\be
{\bar{\eta}}_{ab} = {\sigma}^{-4} \, \eta_{ab}; \qquad {\bar{\eta}}^{ab} =
{\sigma}^{4} \, \eta^{ab}.
\label{Nmetric}
\ee
As a direct inspection shows, $\bar{\eta}_{ab}$ is in fact invariant under the
conformal Poincar\'e group ${\mathcal Q}$. Therefore, if $ds^2 = \eta_{ab}
\, dx^a dx^b$ is the Minkowski interval, the corresponding cone-space  ``conformal
interval'' will be
\be
d\bar{s}^2 = {\bar{\eta}}_{ab} \, d{x}^a d{x}^b.
\label{confin}
\ee
It is worthy mentioning finally that, in this new (maxi\-mal\-ly--symmetric) spacetime
physics will be quite unusual: the ordinary notions of space and time do not exist,
there is no place for the usual concept of movement, no Planck length can be defined,
and so on \cite{aap}.
\end{appendix}


\begin{thebibliography}{99}
\bibitem{sign}
We use a metric tensor with signature $(+1, -1, -1,  -1)$.
\bibitem{spdep}
A space-dependent cosmological term would lead to a non-isotropic space, and will not
be considered.
\bibitem{hands}
Although put by hands, solution (\ref{maden}) is quite reasonable. Notice, for
example, that for a constant cosmological term, it reduces to the well known solution
with $n=3 \omega_m$ of the standard FRW model.
\bibitem{foot4}
A non-vanishing integration constant would correspond simply to a different choice for
the origin of time, and can consequently be ignored.
\bibitem{foot5}
Of course, the proportional  constant will depend on the parameters $n$ and $\omega_m$.
\bibitem{foot6}
Notice that, according to our convention, the de Sitter space has negative, whereas the
anti-de Sitter space has positive scalar curvatures. Of course, both of them have
negative Gaussian curvature.
\bibitem{foot8}
Of course, in order to allow a time increasing $\Lambda$, the range of values of $n$
will necessarily be different from that presented in Eq.~(\ref{nrange}), and used in
section 4.
\end{thebibliography}
\end{document}